\documentstyle{l-aa}

\begin{document}
\title { Disk$-$jet connection in GRS 1915$+$105: X-ray soft dips as cause
of radio flares} 
\author{S. Naik and A.R. Rao }
\institute { Tata Institute of Fundamental Research, Homi Bhabha Road, Mumbai, 400 005, India }
\offprints { S. Naik  {\it snaik@cygnus.tifr.res.in, sachi@tifr.res.in}}
\thesaurus{02.01.2; 02.02.1; 13.25.5; 08.09.2 GRS 1915+105} 
\maketitle
\markboth{Naik and Rao: Disk-Jet connection in GRS 1915+105}{ }

\maketitle

\begin{abstract}

   We have examined the radio emission characteristics of the micro-quasar
GRS~1915$+$105 during its X-ray emission states  as classified by Belloni et al.
(2000). We find that the radio emission is high during the $\chi_1$ and 
$\chi_3$ states (the radio ``plateau'' state) and also during the
$\beta$ and $\theta$ states when X-ray soft dips are present in the 
X-ray light curve. For all the other X-ray states we find that the 
radio emission is low ($<$ 20 mJy at 2.25 GHz). This result supports
the suggestion made by Naik et al. (2000) that the radio flares are
caused by a series of X-ray dip events. To further confirm this result,
we have made a systematic study of all the PCA RXTE observations for those
days when a radio flare was present. We find 11 such observations and 
find that all of them are of class $\beta$, $\theta$ or $\chi$. 
Further, we have classified all the RXTE PCA observations obtained from 1996
November to 2000 February (when the radio data is available) and confirm that
the radio flux, on an average, is much higher during the states $\beta$, 
$\theta$ and $\chi$. Based on these results, we argue that the
radio flares are caused by the X-ray dips in the source.

\keywords{ accretion, accretion disks - black hole physics - X-rays: stars - stars: individual - GRS 1915+105}  
\end{abstract}

\medskip

\section{Introduction}

The Galactic micro-quasar GRS~1915+105 was discovered
in 1992 with the WATCH instrument on-board the GRANAT satellite
(Castro-Tirado et al. 1992). Subsequent radio observations
of the source showed the presence of first superluminal radio source 
at a distance of 12.5 $\pm$ 1.5 kpc in our Galaxy (Mirabel \& Rodriguez 1994). 
The source is beleived to be a black hole because of the similarities in the 
spectral and temporal properties with the other superluminal Galactic radio 
source GRO J1655$-$40 (Zhang et al. 1996).

GRS~1915+105 is a bright X-ray source, emitting at a luminosity
of more than 10$^{39}$ erg s$^{-1}$ for extended periods.
Observations with BATSE on CGRO 
(Harmon et al. 1994) and with SIGMA on Granat (Finoguenov et al. 1994) 
showed the highly variable nature of the source in the hard X-rays. 
QPOs are seen in the power density spectra of the source in the frequency 
range of 0.001 $-$ 64 Hz (Morgan et al. 1997). Chen et al. 
(1997) found that the narrow QPO emission is a characteristic 
property of the hard branch which is absent in the soft-branch. Muno, 
Morgan \& Remillard (1999) found that the low-frequency QPOs (0.5 $-$ 10 Hz) 
can be used as a tracer of the spectral state of the source. 

From radio observations, Mirabel \& Rodriguez (1994) found that 
GRS 1915$+$105 produces double-sided relativistic ejections of plasma
clouds which appears to have superluminal motions.  Pooley \&
Fender (1997) have done a systematic monitoring of the source at 15 GHz
with the Ryle Telescope and found periodic oscillations in the range 
20 $-$ 40 minutes. From simultaneous observations at 
different wavelengths, the correlation between the X-ray variability and 
the emission at radio and infrared wavelengths is established (Pooley \& 
Fender 1997; Fender et al. 1997; Fender \& Pooley 1998; Eikenberry et al. 
1998; Mirabel et al. 1998; Fender et al. 1999). They showed that 
the observed multi-wavelength  behavior is consistent with the scenario 
of plasma ejection from the instabilities in the inner accretion disk of 
the black hole which expands away from the source adiabatically producing 
radio jets.

 Recently, Naik et al. (2000) have reported the detection of a series
of X-ray dips during a huge radio flare of strength 0.48 Jy (at 2.25 GHz).
They argue that a large number of such X-ray dips can account for the
radio flare emission. To investigate whether such X-ray dips are indeed
present during other radio flares, in this paper we present a
detailed analysis of the association of X-ray intensity states with 
the radio emission. We use the classification of X-ray intensity states
suggested by Belloni et al. (2000) and find that the high radio emission
is uniquely identified with only 3 X-ray intensity states: the 
radio-loud hard state or the ``plateau'' state (see Fender et al. 1999)
and the $\beta$ and $\theta$ states, when X-ray dips are present.

\section{Analysis and Results }

 Belloni et al. (2000) have classified all the publicly available 
RXTE/PCA observations from 1996 January to 1997 December into 12 different 
classes. The classification is based on the structure of the X-ray light 
curve and the nature of the color diagram. 
The X-ray variability characteristics range from steady emission for long
durations (classes $\phi$ and $\chi$), short period flickering variability at
different amplitudes (classes $\gamma$, $\mu$ and $\delta$), large amplitude
variations called inner disk oscillations (Belloni et al. 1997) or
regular/irregular bursts (Yadav et al. 1999) where the intensity and
hardness ratio are anti-correlated (classes $\lambda$, $\kappa$ and $\rho$),
and large amplitude variations accompanied by soft dips (classes
$\theta$ and $\beta$). The remaining two classes can be treated as variants
of the above: class $\alpha$ as source repeatedly moving between
class $\chi$ and $\rho$, class $\nu$ as class $\beta$ with much shorter soft
dips. 

To understand the disk-jet 
connection in the source, we have collected the radio flux density for these
12 different X-ray intensity classes at 2.25 GHz and 8.3 GHz from the 
NSF-NRAO-NASA Green Bank Interferometer public domain data. To determine
the radio flux density for each class and to compensate the observed
delay in X-ray and radio wavelengths (Mirabel et al. 1998), we have selected
the radio data in the time range of 2 hours earlier and 6 hours
later to the starting time of the X-ray data from RXTE/PCA and taken the
average value of the radio flux density (if more than one observation exists) 
over the selected time range for 
each observations. Finally the flux densities of each 
observation in a class were averaged out to get the flux density for that 
class. Out of 163 observations, analyzed by Belloni et al., 
there are 89 RXTE/PCA observations after 1996 November 22,
the starting of GBI radio data. As the radio data are not continuous, 
out of these 89 observations, radio data are 
available only for 44 observations in the above selected time range. 
However, for observations in classes $\theta$, $\nu$, $\phi$, $\chi1$ 
and $\chi3$, GBI radio data are not available in the selected time range.

\begin{table*}[]
\caption{Available radio flux during the PCA observations of different classes}
\begin{flushleft}
\begin{tabular}{llllllll}
\hline
\hline
Class  &No. of Obs &Radio flux &at 2.25 &GHz (Jy) &Radio flux &at  8.3 &GHz (Jy)\\
       &           &Avg. &Min &Max                 &Avg. &Min &Max \\	
\hline
\hline
$\alpha$  &7	 &0.014 $\pm$ 0.002 &0.007   &0.027   &0.015 $\pm$ 0.003   &0.007  &0.031\\
$\beta$	  &9	 &0.119	$\pm$ 0.052 &0.008   &0.342   &0.076 $\pm$ 0.027   &0.008  &0.185\\
$\gamma$  &5	 &0.009 $\pm$ 0.0005 &0.007   &0.010  &0.008 $\pm$ 0.0008  &0.006  &0.010\\
$\delta$  &3	 &0.010	$\pm$ 0.0025 &0.007   &0.015   &0.007 $\pm$ 0.001 &0.005   &0.009\\
$\kappa$  &1	 &0.004	 &0.004	   &-----    &0.013	&0.013	 &-----\\	
$\lambda$ &1	 &0.012  &0.012    &-----    &0.013	&0.013	 &-----\\
$\mu$	  &3	 &0.008 $\pm$ 0.0001 &0.008    &0.008    &0.011 $\pm$ 0.003 &0.007  &0.019\\
$\rho$	  &6	 &0.008 $\pm$ 0.0007 &0.005    &0.010    &0.008 $\pm$ 0.0005 &0.006 &0.009\\
$\chi$2	  &3	 &0.013 $\pm$ 0.0057 &0.007    &0.025    &0.007 $\pm$ 0.0024 &0.004 &0.012\\
$\chi$4	  &7	 &0.015	$\pm$ 0.0033 &0.007    &0.027    &0.016 $\pm$ 0.0043 &0.006 &0.032\\ 
\hline
\hline
\end{tabular}
\end{flushleft}
\end{table*}

From the RXTE/ASM light curve of the source (Fig. 9, Belloni et al. 2000),
the subclasses $\chi$1, $\chi$2 and $\chi$3 are extended and well separated.
Although the radio data do not exist for the PCA observations in the
selected time range for $\chi$3, the flux density is taken from the GBI 
data in the time range as shown in the Fig. 9 of Belloni et al. (2000).
The flux density for classes $\theta$, $\nu$, $\phi$ and $\chi$1 were 
taken from Fig. 1 of Pooley \& Fender (1997). To get the 
flux density at 8.3 GHz corresponding to the values at 15 GHz in Fig. 1 by 
Pooley \& Fender (1997), we have used the relation given
by Mirabel et al. (1998). The flux density at 2.25 GHz and 8.3 GHz for
all the 12 classes is shown in Fig. 1. The upper panel shows the average 
flux density at 2.25 GHz and the lower panel shows at 8.3 GHz for different
classes. The classes for which the data are taken from Pooley 
\& Fender (1997) are marked by an asterisk in the figure to identify from 
the others. The details of the radio observations are given in Table 1.
Out of 15 different classes of PCA observations (including 4 sub-classes
of $\chi$), the GBI radio data are available
for 12 classes of observations in the above selected time range.
From the table, it is seen that the average radio flux densities are 
$\sim$ 120 mJy at 2.25 GHz and $\sim$ 80 mJy at 8.3 GHz during the PCA observations 
of class $\beta$. The observed maximum and minimum radio flux density during the PCA 
observations of class $\beta$ are found to be 342 mJy and 8 mJy at 2.25 GHz and 185 
mJy and 8 mJy at 8.3 GHz respectively. Out of 9 PCA observations of class $\beta$, 
during 6 observations, the average flux densities at 8.3 GHz are found to be more than 37 mJy 
with a  maximum of 205 mJy and during the other 3 observations, the flux densities are 
$\simeq$ 10 mJy. From a careful analysis of these PCA observations of class $\beta$ with 
low value of average flux densities, it is seen that the structure of the light curve of 
only one orbit of each of 2 RXTE/PCA observations are of class $\beta$. For the 
third PCA observation
with low flux density, it is seen that the GBI radio data are not available for about 15 hours
after the begining of the X-ray soft dip in the light curve which is supposed to be the cause
of the radio flares followed by a flux density of $\sim$ 35 mJy at 2.25 GHz. 
From the table, however, it is seen that the average flux densities at 2.25 GHz during the 
PCA observations of other classes are $\leq$ 15 mJy. Hence, we can conclude that the radio flux 
density during the RXTE/PCA observations of class $\beta$ are high in comparison to the other 
observations.

\begin{figure}[t]
\vskip 6.0 cm
\includegraphics{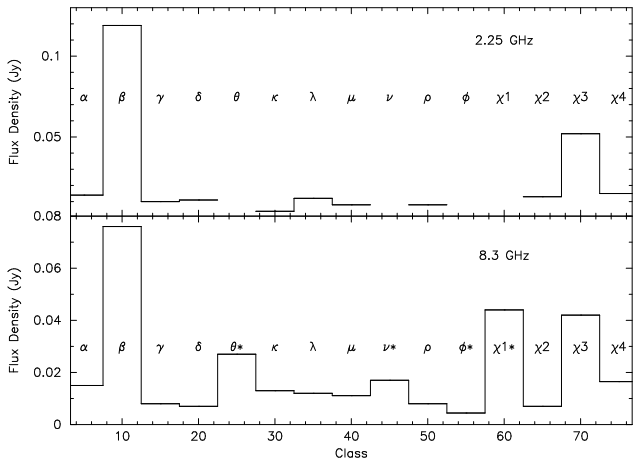}
\caption {  The average radio flux for GRS 1915$+$105 at 2.25 GHz (upper panel)
and 8.3 GHz (lower panel) for all the twelve classes of X-ray observations, 
classified by Belloni et al. (2000) during the period 1996 June to 1997 
December. The radio data is taken from  NSF-NRAO-NASA Green Bank 
Interferometer public domain data. The data for classes marked by  an 
asterisk are derived from the 15 GHz data (Pooley \& Fender 1997).}
\end{figure}

From the figure, it is seen that the flux densities for classes $\beta$, 
$\theta$, $\chi$1 and $\chi$3, are higher. It is seen that
there was a huge radio flare during MJD 50751 to 50753 during which the
average flux density was more than 300 mJy at 2.25 GHz. All the RXTE/PCA 
observations (3) during this period are of class $\beta$. 
Apart from this huge radio flare, there were several small flares during
the X-ray observations of class $\beta$. Although in Belloni et al. (2000)
classification, there are less number of observations of class $\theta$, 
a radio flare at 15 GHz was observed (Fig. 1 and 4, Pooley \& Fender 1997) 
during the period of PCA observations in class $\theta$ (MJD 50250 $-$ 50254).
During class $\chi$, the flux density was steady without any evidence  
for oscillations and described as ``plateau'' by Fender et al. (1999).  
From these results, we suggest that the source is in high radio emission 
state during the X-ray classes $\beta$, $\theta$ and $\chi$.
    
We have verified the above result by considering X-ray data only for 
radio flares. For this purpose, we define a radio flare as occasions
when the flux density stayed above 0.1 Jy at 2.25 GHz for more than 
half a day. We find a total number of 46 such radio flares in the time 
range 1996
November 22 to 2000 February 5 and tried to find out the RXTE/PCA observations 
during these selected radio flares. Out of these 46 radio flares, X-ray data
is available for 10 flares. The X-ray light curve during 4 flares 
are of type $\theta$ (3) and $\beta$ (1), during 5 radio flares, the light 
curves are of $\chi$ type. However the X-ray light curve for one radio 
flare that occuered during MJD 50923.0 $-$ 50923.8 can not be classified
under any class. From the structure of the light curve, it appears 
to be of class $\theta$. There were no RXTE/PCA observations during 
the remaining 
36 radio flares. The occupation time of these selected flares is found to
be 90 days out of which 3 orbits of RXTE/PCA observations ($\sim$ 40 $-$ 50
minutes each) contain soft dips of classes $\beta$, 4 orbits contain soft 
dips of class $\theta$ and 26 orbits are of class $\chi$.  
Fig. 2 shows the radio light curve at 2.25 GHz with GBI 
data. The presence of RXTE/PCA observations of class $\theta$ and $\beta$ 
during the selected radio flares with flux density more than 0.1 Jy are 
indicated by arrows. The RXTE/PCA observations of class $\chi$ during the 
4 radio flares and the other one are shown in the insets of Fig. 2. From 
these figures, the RXTE/PCA observations are found to be towards the end 
of the radio flare except for one event during MJD 51003.0 $-$ 51005.4 when the 
X-ray light curves are in the class $\chi$. This strongly suggest that 
the radio flaring is closely associated with the occurance of X-ray states 
$\theta$ and $\beta$ and the high and steady radio emission state, called 
as ``plateau'' is associated with the low-hard X-ray state.

\begin{figure}[t]
\vskip 6.0 cm
\includegraphics{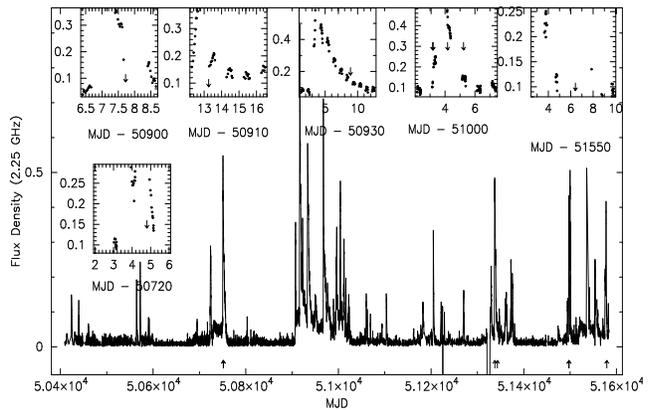}
\caption {  The presence of soft dips in the X-ray light curve of
GRS 1915$+$105 from RXTE/PCA during the huge radio flares observed 
with GBI is marked with arrows. The RXTE/PCA observations when class $\chi$ 
was was present (see text) are shown as arrows in the insets.} 
\end{figure}

In Fig. 3, we have plotted the light curves and hardness
ratio (count rate in 5 $-$ 13 keV energy range/count rate in 2$-$5
keV energy range) of one observation from each classes of $\theta$
and $\beta$. The similarity in both the classes is the presence of 
soft X-ray dip in the light curve with identical properties i.e.
soft spectrum (hardness ratio), absence of the low frequency QPO 
(refer Fig. 1e and 1f of Muno et al. 1999). These soft dips of period
of $\sim$ 100 s are seen only in the classes $\theta$ and $\beta$ out
of all 12 classes of Belloni et al. (2000). The presence of soft
Dips in the X-ray light curve, associated with the radio flaring,
strongly supports the suggestion given by Naik et al. (2000)
that the evacuation of matter from the accretion disk during the presence 
of soft dips in the X-ray light curve produces the radio flares.

\begin{figure}[h]
\vskip 6.0 cm
\includegraphics{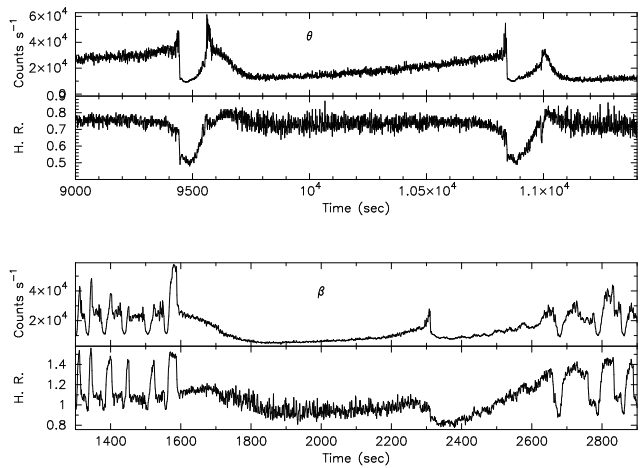}
\caption {  The X-ray light curve and hardness ratio (H. R.) of GRS 1915$+$105 
obtained from RXTE/PCA observations for the classes $\theta$ (1999 June 8)  
and $\beta$ (1997 August 31). X-ray dips with low hardness ratio is seen in
both the cases.}
\end{figure}

We have analyzed all the publicly available data from RXTE/PCA from 1996 
November 22 to 2000 February 05 when the GBI radio is available. Out of 
333 number of RXTE/PCA observations, there are  17 observations of class
$\beta$ and 20 observations of class $\theta$. The remaining 297 
observations pertain to the other classes of Belloni et al. (2000) 
classification. However, the radio flux densities are available only
for 174 PCA observations in the above selected time range. The details 
of the analysis are given in Table 2.

\begin{table*}[t]
\caption{Average radio flux density at 2.25 GHz during the PCA observations 
of classes $\beta$, $\theta$, $\chi$ and the rest during 1999 November 22 to 2000 February 5}
\begin{flushleft}
\begin{tabular}{llllllll}
\hline
\hline
Class  &No. of Obs. &Flux density &at 2.25 &GHz(Jy) &Flux density &at 8.3 &GHz(Jy)\\
       &           &Avg.       &Max 	 &Min    &Avg.     &Max    &Min   \\	
\hline
\hline
$\beta$	  &10	 &0.128  $\pm$ 0.046	&0.347	&0.008    &0.080 $\pm$ 0.023  &0.186   &0.008\\
$\theta$  &14	 &0.041	 $\pm$ 0.016    &0.187	&0.008    &0.032 $\pm$ 0.005  &0.074   &0.013\\
$\chi1$/$\chi3$	  &35	 &0.096  $\pm$ 0.014 	&0.226	&0.009    &0.058 $\pm$ 0.0062  &0.148   &0.010\\
$\chi2$/$\chi4$	  &20	 &0.019  $\pm$ 0.003 	&0.054	&0.006    &0.016 $\pm$ 0.002  &0.032   &0.004\\
others	  &95	 &0.013	 $\pm$ 0.001	&0.048	&0.005    &0.012 $\pm$ 0.001  &0.048   &0.004\\ 
\hline
\hline
\end{tabular}
\end{flushleft}
\end{table*}

Out of the 17 observations of class $\beta$ and 20 
observations of class $\theta$, the radio flux densities are available 
only for 10 and 14 observations with the average flux densities of 0.128 Jy 
and 0.041 Jy at 2.25 GHz and 0.080 Jy and 0.032 Jy at 8.3 GHz respectively in 
the above selected time range. Similarly out of 297 RXTE/PCA
observations of the classes other than $\beta$ and $\theta$, the radio
data are available for 55 observations of class $\chi$. As it is seen from
Fig. 1 that the radio flux densities during the PCA observations of subclasses 
$\chi1$ and $\chi3$ are higher than subclasses  $\chi2$ and $\chi4$, we divided
these 55 observations of class $\chi$ into 2 subclasses (the PCA observations in
subclasses $\chi1$ and $\chi3$ into one class and observations in subclasses $\chi2$ 
and $\chi4$ into the other class) based on the RXTE/ASM count rate. 
The RXTE/ASM count rate for the source during the PCA observations of subclasses 
$\chi2$ and $\chi4$ is found to be $\leq$ 30 ASM counts whereas during
subclasses $\chi1$ and $\chi3$, it is $\geq$ 35 ASM counts. 
It is found that the average radio flux density at 2.25 GHz for the class $\chi1$/$\chi3$
is 0.096 Jy for 35 PCA observations, whereas it is 0.019 Jy
for the other class $\chi2$/$\chi4$ for 20 PCA observations.
During the 95 PCA observations of other classes, the average flux density at 2.25 GHz 
is found to be 0.013 mJy. From this analysis, it is verified that the source
remains in a high radio state during the X-ray states of class $\beta$, $\theta$ and
$\chi1$/$\chi3$. During the X-ray state of class $\chi1$/$\chi3$, the source shows a steady 
and flat spectrum with little oscillations (Fender et al. 1999) in radio bands.
Hence, it is verified that radio flares occur mainly in 
$\beta$ and $\theta$ classes of X-ray states.

\section{Discussion and conclusions}

The micro-quasar GRS 1915$+$105 offers a unique opportunity to study the
connection between the observed radio jets and the accretion disks which
is thought to be present in distant Quasars. Attempts have been made to
relate the changes in X-ray emission from the accretion disk with the
amount of mater ejected from the system. 
Belloni et al. (1997) discovered
a series of outbursts during which the change in state of the source occurs
within $\sim$ 10 $-$ 100 seconds and described the change as the appearance 
and disappearance of the inner accretion disk. Yadav 
et al. (2000) studied the same burst events and described
the two intensity states as the low-hard and high-soft state of the source.
The radio flux density during these periods was found be low ($\sim$ 8 mJy 
at 2.25 GHz). There are certain periods when both the observed flux in X-ray 
and radio are low and are described as radio-quite hard state (Muno et al. 
1999). The periods when both the X-ray and radio fluxes are high without
the evidence of any oscillations, are called as radio-loud hard-steady state.
Fender et al. (1999) described this class as the ``plateau'' state.
Pooley \& Fender (1997) speculated that this plateau state may correspond 
to the major radio ejections in the source. 

Class $\beta$ is a peculiar type of X-ray emission associated with the change 
from a high oscillating state to a low-hard state with the presence 
of a soft dip during the recovery with the characteristics of low
intensity and absence of low frequency QPO, followed by a gradual
return to the high oscillating state (Belloni et al. 2000). 
These observations are associated with the synchrotron flares in 
radio (Mirabel et al. 1998; Fender \& Pooley 1998) and infrared 
(Eikenberry et al. 1998). From simultaneous X-ray and infrared 
observations, Eikenberry et al. (1998) made a strong argument that 
the onset of radio/infrared flare is associated with the soft dip 
rather than the gradual change to the low-hard state. 
Eikenberry et al. (2000) identified a series of soft X-ray 
dips (class $\theta$, 
Belloni et al. 2000) identical with the dips seen during 
the synchrotron infrared flares, coincident with  faint infrared flares.
Radio oscillations on a time scale of
20 $-$ 30 minutes are seen to be accompanied by a series of soft X-ray
dips (Fig. 10, Dhawan et al. 2000), similar to the oscillations observed
by Fender et al. (1999). Since the soft dip events are associated with
the jet emission, Naik et al. (2000) proposed that the huge 
radio flares are produced by a series of such soft X-ray dips. 

  It was known earlier that the hard steady state of GRS 1915+105
showed radio emission (the $\chi1$/$\chi3$ state or the `plateau' state) with
flat spectrum radio emission. It was also know earlier that $\beta$ and
$\theta$ states are also associated with radio flares, the so called
`baby jets' (Mirabel et al. 1998; Eikenberry et al. 1998, 2000). It was, 
however, not known very
conclusively what accretion disk characteristics (as observed in X-rays)
are associated with steep spectrum radio flares, some of which are
observed to be associated with superluminal jet emission. Fender et al. (1999)
observed radio oscillations during the start of superluminal ejections
and they speculated that the inner disk oscillations give rise to
radio oscilations. The present work shows conclusively that the inner
disk oscillations (classes $\lambda$, $\kappa$ and $\rho$) are not associated
with radio emission. Naik et al. have pointed out that the source was
exhibiting such inner disk oscillations for about a month in 1997
June (Yadav et al. 1999) when the radio flux was consistently below 20 mJy.

 Naik et al. (2000) have, for the first time, monitored the source in 
sufficient 
time resolution (1 s) for long duty cycles (a few thousand seconds per
day, continuously for several days) simultaneously with a radio flare. They
detected a series of soft dips (class $\theta$) which are distinctly different
from the hard dips of inner disk oscillations. They have speculated that
a series of X-ray dips is responsible for the radio flare. Hence the
following scenario emerges from the present study: 1) the steady state
$\chi$1 and $\chi$3 can give flat spectrum radio source associated with 
compact ($<$ 10 AU) radio core, 2) The soft X-ray dips can generate small 
size jet emission and 3) a large collection of X-ray dips causes
the radio core to be detached and move out at a super-luminal speed.

We must, however, point out that the association between radio emission
and the dip events is of statistical nature. Hence we critically
examine below the evidence for this association. 
If the onset of a huge radio flare (signifying the emission
of a superluminal ejecta) is associated with an X-ray emission
characteristic observed so far, it has to be necessarily the
X-ray dip events. It is, however, quite possible that the X-ray
dips do not produce the radio flare due to the following reasons:
a) onset of a radio flare does not produce any observable X-ray emission
characteristics  or b) onset of a radio emission produces a new type of
X-ray emission characteristic which has not been observed so far   and 
c) the X-ray dip events could be the manifestations of 
a disturbed accretion disk, after the production of an ejecta.
We argue here that none of these scenarios are likely.

 Fender et al. (1999) have worked back the onset time of the superluminal blobs
and found that the source was showing 20 - 30 minute radio oscillations.
Such a short time scale should be arising from regions close to the compact 
object and hence it is very unlikely that the X-ray emission characteristics
will remain unaffected.  It should be noted here that radio oscillations
at similar periods are observed along with a series of X-ray dip events
(see Fig 10. of Dhawan et al. 2000). The radio ejecta contains
energy of $\sim$ 10$^{43}$ $-$ 10$^{44}$ ergs and hence the accretion 
disk event
responsible for it should last for about 12 hours (see Fender et al. 1999).
Since there are 46 radio flares (above 0.1 Jy) during the GBI monitoring
period, it is extremely unlikely that such an event has gone unnoticed so far.
Hence we can conclude that presence of a collections of X-ray dips are responsible for 
the radio flare and consequently the jet emission.
A continuous X-ray monitoring during a radio flare should clarify
these questions.

\begin{acknowledgements}

We thank the members of the RXTE and NSF-NRAO-NASA GBI team for making the
data publicly available. The Green Bank Interferometer is
a facility of the National Science Foundation operated by the NRAO in
support of NASA High Energy Astrophysics programs. We thank P. C. Agrawal
for his useful comments.

\end{acknowledgements}

{}




\end{document}